\documentclass[amsmath,amssymb,aps,prx,reprint,superscriptaddress,onecolumn,longbibliography]{revtex4-2}

\usepackage[utf8]{inputenc}
\usepackage{geometry}
\usepackage{graphicx}
\usepackage{fancyhdr}
\usepackage[usenames,dvipsnames,svgnames]{xcolor}
\usepackage{tcolorbox}
\usepackage{orcidlink} 
\usepackage{titlesec}
\usepackage{setspace} 
\usepackage{amsmath}
\usepackage{amsfonts,amssymb}
\usepackage{mhchem}
\usepackage{needspace}
\usepackage{siunitx}

\definecolor{morandiblue}{RGB}{104, 131, 155}     
\definecolor{morandidark}{RGB}{54, 84, 110}       
\definecolor{morandigreen}{RGB}{126, 142, 118}    
\definecolor{brightcerulean}{RGB}{58, 144, 196}  

\makeatletter
\def\@date{}
\def\frontmatter@date{}
\def\frontmatter@authorformat{\normalsize\normalfont\centering}
\def\frontmatter@above@affiliation{\vspace*{-1.2ex}} 

\def\cat@comma@active{\catcode`\,12}%

\let\GPUMDkit@orig@frontmatter@author@produce\frontmatter@author@produce
\def\frontmatter@author@produce{%
  \begingroup
    \hsize=0.98\textwidth\relax         
    \linewidth=0.98\textwidth\relax
    \tiny                    
    \raggedright
    \sloppy
    \setlength{\parskip}{0.5pt}    
    \setlength{\parindent}{0pt}
    \GPUMDkit@orig@frontmatter@author@produce
  \endgroup
}
\makeatother

\usepackage{hyperref}
\hypersetup{
    colorlinks=true,
    linkcolor=morandidark, 
    citecolor=DarkBlue, 
    urlcolor=DarkBlue,
}

\titleformat*{\section}{\large\bfseries\color{morandidark}}
\titleformat*{\subsection}{\large\bfseries\color{morandidark}}
\titleformat*{\subsubsection}{\bfseries\color{morandidark}}

\renewcommand{\headrulewidth}{0.8pt}
\renewcommand{\headrule}{\hbox to\headwidth{\color{morandidark}\leaders\hrule height \headrulewidth\hfill}}

\fancypagestyle{titlepage}{
    \fancyhf{}
    \renewcommand{\headrulewidth}{0.8pt}
    \renewcommand{\headrule}{\hbox to\headwidth{\color{morandidark}\leaders\hrule height \headrulewidth\hfill}}
    \fancyhead[L]{\includegraphics[height=0.8cm]{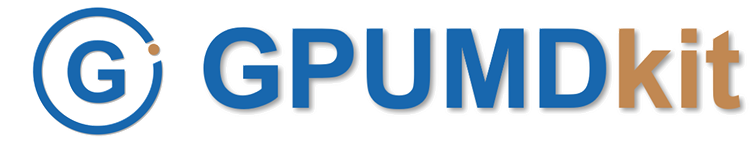}}
    \fancyhead[R]{}
	\vspace{0.2cm}
}

\begin{document}

\title{\textbf{\color{morandidark}\large GPUMDkit: A User-Friendly Toolkit for GPUMD and NEP}\vspace{0.3cm}}

\author{Zihan Yan\,\orcidlink{0000-0002-8911-6549}}
\email[]{yanzihan@westlake.edu.cn}
\affiliation{Department of Materials Science and Engineering, School of Engineering, Westlake University, Hangzhou, Zhejiang 310030, China}
\affiliation{School of Materials Science and Engineering, Zhejiang University, Hangzhou, Zhejiang 310027, China}

\author{Denan Li}
\affiliation{Department of Physics, School of Science, Westlake University, Hangzhou, Zhejiang 310030, China}

\author{Xin Wu\,\orcidlink{0000-0002-7179-371X}}
\affiliation{Institute of Industrial Science, The University of Tokyo, Tokyo 153-8505, Japan}

\author{Zhoulin Liu\,\orcidlink{0000-0001-8755-082X}}
\affiliation{School of Science, Harbin Institute of Technology, Shenzhen 518055, Guangdong, China}

\author{Chen Hua\,\orcidlink{0009-0008-8502-1068}}
\affiliation{State Key Lab of Cryogenic Science and Technology, Technical Institute of Physics and Chemistry, Chinese Academy of Sciences, Beijing 100190, China}
\affiliation{School of Future Technology, University of Chinese Academy of Sciences, Beijing 100049, China}

\author{Boyi Situ\,\orcidlink{0000-0002-2275-7165}}
\affiliation{Department of Materials Science and Engineering, School of Engineering, Westlake University, Hangzhou, Zhejiang 310030, China}
\affiliation{School of Materials Science and Engineering, Zhejiang University, Hangzhou, Zhejiang 310027, China}

\author{Hao Yang}
\affiliation{School of Physical Science and Technology, ShanghaiTech University, Shanghai 201210, China}

\author{Shengjie Tang}
\affiliation{Department of Materials Science and Engineering, School of Engineering, Westlake University, Hangzhou, Zhejiang 310030, China}
\affiliation{School of Materials Science and Engineering, Zhejiang University, Hangzhou, Zhejiang 310027, China}

\author{Benrui Tang}
\affiliation{College of Physical Science and Technology, Bohai University, Jinzhou 121013, China}


\author{Ziyang Wang}
\affiliation{School of Energy and Power Engineering, Key Lab of Ocean Energy Utilization and Conservation of Ministry Education, Dalian University of Technology, Dalian, 116024, China}

\author{Shangzhao Yi}
\affiliation{Suzhou Institute of Nano-Tech and Nano-Bionics, Chinese Academy of Sciences, Suzhou 215123, China}

\author{Huan Wang\,\orcidlink{0000-0002-4931-3898}}
\affiliation{School of Materials Science and Engineering, Wuhan University of Technology, Wuhan 430070, China}

\author{Dian Huang}
\affiliation{MOE Key Laboratory of Thermo-Fluid Science and Engineering, School of Energy and Power Engineering, Xi'an Jiaotong University, Xi'an 710049, China}


\author{\hbox{Ke Li}\,\orcidlink{0009-0004-5972-2072}}
\affiliation{Department of Thermal Science and Energy Engineering, School of Engineering Science, University of Science and Technology of China, Hefei, Anhui 230027, China}

\author{Qilin Guo}
\affiliation{Hangzhou International Innovation Institute, Beihang University, Hangzhou 311115, China}

\author{Zherui Chen}
\affiliation{Future Technology School, Shenzhen Technology University, Shenzhen 518118, China}

\author{Ke Xu}
\affiliation{College of Physical Science and Technology, Bohai University, Jinzhou 121013, China}

\author{Yanzhou Wang}
\affiliation{School of Electronic Engineering, Chengdu Technological University, Chengdu 611730, China}

\author{Ziliang Wang\,\orcidlink{0009-0004-9310-1501}}
\affiliation{National Engineering Laboratory for Reducing Emissions from Coal Combustion, Shandong Key Laboratory of Green Thermal Power and Carbon Reduction, Shandong University, Jinan, Shandong 250061, China}

\author{Gang Tang}
\affiliation{School of Interdisciplinary Science, Beijing Institute of Technology, Beijing 100081, China}

\author{\hbox{Shi Liu}}
\affiliation{Department of Physics, School of Science, Westlake University, Hangzhou, Zhejiang 310030, China}

\author{Zheyong Fan\,\orcidlink{0000-0002-2253-8210}}
\affiliation{College of Physical Science and Technology, Bohai University, Jinzhou 121013, China}

\author{Yizhou Zhu\,\orcidlink{0000-0002-5819-7657}}
\email[]{zhuyizhou@westlake.edu.cn}
\affiliation{Department of Materials Science and Engineering, School of Engineering, Westlake University, Hangzhou, Zhejiang 310030, China}

\maketitle


\vspace{-0.5cm}
\begin{center}
\begin{tcolorbox}[colback=morandiblue!8, colframe=morandiblue, boxrule=1pt, arc=4pt, width=0.95\textwidth]
\begin{center}
\textbf{\textcolor{morandiblue}{\large Abstract}}
\end{center}
\vspace{0.2cm}

Machine-learned interatomic potentials have revolutionized molecular dynamics simulations by providing quantum-mechanical accuracy at empirical-potential speeds. The graphics processing unit molecular dynamics (GPUMD) package, featuring the highly efficient neuroevolution potential (NEP) framework, has emerged as a powerful tool in this domain. However, the complexity of force field development, active learning, and trajectory post-processing often requires extensive manual scripting, imposing a steep learning curve on new users. To address this, we present GPUMDkit, a comprehensive and user-friendly toolkit that streamlines the entire simulation workflow for GPUMD and NEP. GPUMDkit integrates a suite of essential functionalities, including format conversion, structure sampling, property calculation, and data visualization, accessible through both interactive and command-line interfaces. Its modular, extensible architecture ensures accessibility for users of all experience levels while allowing seamless integration of new features. By automating complex tasks and enhancing productivity, GPUMDkit substantially lowers the barrier to using GPUMD and NEP programs. This article describes the program architecture and demonstrates its capabilities through practical applications.

\vspace{0.2cm}
\noindent \textbf{\small {GitHub Repository:}} \small \url{https://github.com/zhyan0603/GPUMDkit}
\end{tcolorbox}
\end{center}

\clearpage

\section{Introduction}

Machine-learned interatomic potentials (MLIPs) have fundamentally transformed the landscape of atomistic simulations in recent years, bridging the accuracy-efficiency gap that has long constrained molecular dynamics (MD) simulations. MLIPs can capture high-dimensional potential energy surfaces with near-quantum-mechanical accuracy, at a computational cost several orders of magnitude lower than first-principles methods such as density functional theory (DFT). This paradigm shift has enabled machine-learning molecular dynamics (MLMD) simulations of unprecedented scale and duration, unlocking new possibilities for studying various properties of materials, such as mechanical properties~\cite{qi2024interfacial,liu2025predicting}, thermal transport~\cite{xu2023accurate,yu2024dynamic,ying2025highly}, phase transitions~\cite{fransson2023phase,yan2024impact,li2024revealing}, various dynamic phenomena~\cite{li2024revealing,zeng2023molecular}, and more~\cite{zhao2023development,zhao2024general}.

The concept of modern MLIPs was established by Behler and Parrinello in 2007~\cite{behler2007generalized}, who introduced high-dimensional neural network potentials. Subsequent developments have greatly expanded the MLIP landscape. For example, Gaussian approximation potentials employ Gaussian process regression to interpolate potential energy surfaces~\cite{bartok2010gaussian}. Moment tensor potentials offer a polynomial-like potential based on the tensors of inertia of atomistic environments~\cite{novikov2020mlip}. Deep potentials introduced a deep neural network approach that accurately describes local atomic environments and atomic interactions~\cite{zhang2018deep}. More recently, graph neural network-based approaches have gained prominence, with E(3)-equivariant architectures demonstrating exceptional accuracy and data efficiency~\cite{batzner2022}, and the MACE framework further advancing the field through higher-order equivariant message passing~\cite{batatia2022mace,batatia2025design}. Together, these methods offer researchers diverse pathways to accurate and efficient atomistic simulations.

Among them, the neuroevolution potential (NEP) method~\cite{fan2021neuroevolution} based on the separable natural evolution strategy~\cite{schaul2011high} stands out for its unique combination of computational efficiency and accuracy. NEP has demonstrated remarkable success across diverse applications, including large-scale simulations of thermal transport~\cite{dong2024molecular}, radiation damage~\cite{liu2023large}, phase transition~\cite{fransson2023phase,yan2024impact,li2024revealing}, and inorganic and organic systems~\cite{liang2025nep89}. The release of GPUMD 4.0~\cite{xu2025gpumd}, along with the development of NEP89 ~\cite{liang2025nep89}, a universal NEP model across 89 elements, further demonstrates the versatility and accuracy of this approach across diverse material systems, enabling simulations at unprecedented scale on modern GPU architectures.

Despite these advantages, practical use of GPUMD and NEP often demands proficiency with specialized scripts or Python packages. Within the GPUMD ecosystem, several excellent packages have been developed to assist users: NepTrain and NepTrainKit~\cite{chen2025neptrain} provide powerful automated workflows for force field development and enable GUI-based visualization of the training datasets, while Calorine~\cite{lindgren2024calorine} offers a comprehensive Python API designed specifically for various property analysis, and PYSED~\cite{liang2025pysed} is specifically designed to extract kinetic-energy-weighted phonon dispersion and lifetimes from MD simulations. Although these tools form the cornerstone of many advanced workflows, they can still present a barrier for users who prefer an ``out-of-the-box" experience. For researchers focused on specific material systems or properties rather than method development, the most pressing need is a straightforward way to perform standardized tasks through simple command-line operations or intuitive interactive prompts.

To bridge this gap, we developed GPUMDkit. Rather than replacing existing tools, GPUMDkit serves as an integrator that encapsulates a wide range of scripts and modular functionalities into a unified interface, accessible via both command-line and interactive modes. This design allows users to move rapidly from data preparation to property analysis without extensive custom coding, lowering the barrier to using GPUMD and NEP, so users can focus more on scientific questions rather than technical implementation.

\section{The GPUMD Package and the NEP method}

GPUMD is a general-purpose MD simulation package built on a natively optimized CPU + GPU heterogeneous architecture via CUDA/HIP programming~\cite{fan2017efficient,xu2025gpumd}, delivering exceptional computational efficiency for large-scale atomistic simulations. The package supports a wide range of interatomic potentials, including Lennard-Jones~\cite{lennard1931cohesion}, embedded-atom method~\cite{daw1984embedded}, Tersoff~\cite{tersoff1988empirical}, force-constant~\cite{brorsson2022efficient}, deep potential~\cite{zeng2023deepmd}, NEP~\cite{song2024general}, hybrid NEP-ILP~\cite{bu2025accurate} and ILP-SW~\cite{jiang2025moire}. Renowned for its extreme simulation efficiency, it is rapidly gaining popularity among MD packages. The recent release of GPUMD 4.0 has further expanded its capabilities for versatile materials simulations with NEP method~\cite{xu2025gpumd}.

The NEP method was originally proposed by Fan \textit{et al.} in 2021 (NEP1)~\cite{fan2021neuroevolution} and has undergone continuous development through successive generations including NEP2~\cite{fan2022improving}, NEP3~\cite{fan2022gpumd}, and NEP4~\cite{song2024general}, with NEP4 demonstrating improved accuracy for multi-component systems. In the NEP model, the separable natural evolution strategy proposed by Schaul \textit{et al.}~\cite{schaul2011high} is employed to train the neural network potential function, which is why this MLIP is called NEP. The artificial neural network (ANN) model of NEP is a multivariate function, in which the energy is defined as the sum of the atomic energies of all sites:
\begin{equation}
 U_i=\sum_{\mu=1}^{N_{\mathrm{neu}}}{w_\mu^{\left(1\right)}\tanh{\left(\sum_{v=1}^{N_{\mathrm{des}}}{\mathbf{w_{\mu\nu}^{\left(0\right)}}q_v^i}-b_\mu^{\left(0\right)}\right)}}-b^{\left(1\right)}
 \label{nep1}
\end{equation}
Here, the input layer of ANN is  $q_\nu^i$, a high-dimensional descriptor vector, and the output layer is  $U_i$. $\tanh\left(x\right)$ is the activation function of the hidden layer, $N_{\mathrm{neu}}$ is the number of neurons, $N_{\mathrm{des}}$ is the number of components of the descriptor vector, $\mathbf{w_{\mu\nu}^{\left(0\right)}}$ is the matrix of connection weights from the input layer to the hidden layer, $w_\mu^{\left(1\right)}$ is the vector of connection weights from the hidden layer to the output layer, $b_\mu^{\left(0\right)}$ is the bias vector of the hidden layer, and $b^{(1)}$ is the bias of the output layer. 

The descriptors, which consist of high-dimensional vectors that map atomic information into mathematical form, are crucial for ANN models. In the NEP model, the descriptors for atom $i$ consist of a number of radial and angular components. Radial descriptor components are expressed as:
\begin{equation}
  q^i_{n}
  = \sum_{j\neq i} g_{n}(r_{ij})
  \quad\text{with}\quad
  0\leq n\leq n_\mathrm{max}^\mathrm{R}
\label{qin}
\end{equation}
The function $g_n(r_{ij})$ that depends only on the distance between atoms $i$ and $j$ ($r_{ij}$) is called a radial function and has the following expression:
\begin{equation}
  g_n(r_{ij}) = \sum_{k=0}^{N_\mathrm{bas}^\mathrm{R}} c^{ij}_{nk} f_k(r_{ij}),
\label{g_n}
\end{equation}
where
\begin{equation}
  f_k(r_{ij}) = \frac{1}{2}
  \left[
      T_k\left(2\left(r_{ij}/r_\mathrm{c}^\mathrm{R}-1\right)^2-1\right)+1
  \right]
  f_\mathrm{c}(r_{ij}).
\label{f_n}
\end{equation}
Here, $T_k(x)$ represents the $k^{\rm th}$ order Chebyshev polynomial of the first kind, while $f_\mathrm{c}(r_{ij})$ denotes the cutoff function defined as follows:
\begin{equation}
   f_\mathrm{c}(r_{ij}) 
   = \begin{cases}
   \frac{1}{2}\left[
   1 + \cos\left( \pi \frac{r_{ij}}{r_\mathrm{c}^\mathrm{R}} \right) 
   \right],& r_{ij}\leq r_\mathrm{c}^\mathrm{R}; \\
   0, & r_{ij} > r_\mathrm{c}^\mathrm{R}.
   \end{cases}
\label{f_c}
\end{equation}
$n_\mathrm{max}^\mathrm{R}$ and $N_\mathrm{bas}^\mathrm{R}$ in Eq. (\ref{qin}) and Eq. (\ref{g_n}) are tunable hyperparameters in the training process and $r_\mathrm{c}^\mathrm{R}$ in Eq. (\ref{f_n}) and Eq. (\ref{f_c}) is the cutoff distance of the radial descriptor components.

For the angular descriptor components, NEP4 model considers 3-body to 5-body ones. The formulation is similar but not identical to the atomic cluster expansion approach~\cite{drautz2019atomic}. For 3-body angular descriptor components ($0\leq n\leq n_\mathrm{max}^\mathrm{A}, 1\leq l \leq l_\mathrm{max}^\mathrm{3b}$),
\begin{equation}
    q^i_{nl} = \sum_{m=-l}^l (-1)^m A^i_{nlm} A^i_{nl(-m)},
\end{equation}
where
\begin{equation}
    A^i_{nlm} = \sum_{j\neq i} g_n(r_{ij}) Y_{lm}(\theta_{ij},\phi_{ij}),
\end{equation}
and $Y_{lm}(\theta_{ij},\phi_{ij})$ are the spherical harmonics as a function of the polar angle $\theta_{ij}$ and the azimuthal angle $\phi_{ij}$. For more details and the expressions of the 4-body and 5-body descriptor components, we refer to Ref.~\cite{fan2022gpumd}.

\section{GPUMDkit Package}

\begin{figure*}[htb]
  \centering
  \includegraphics[width=0.99\linewidth]{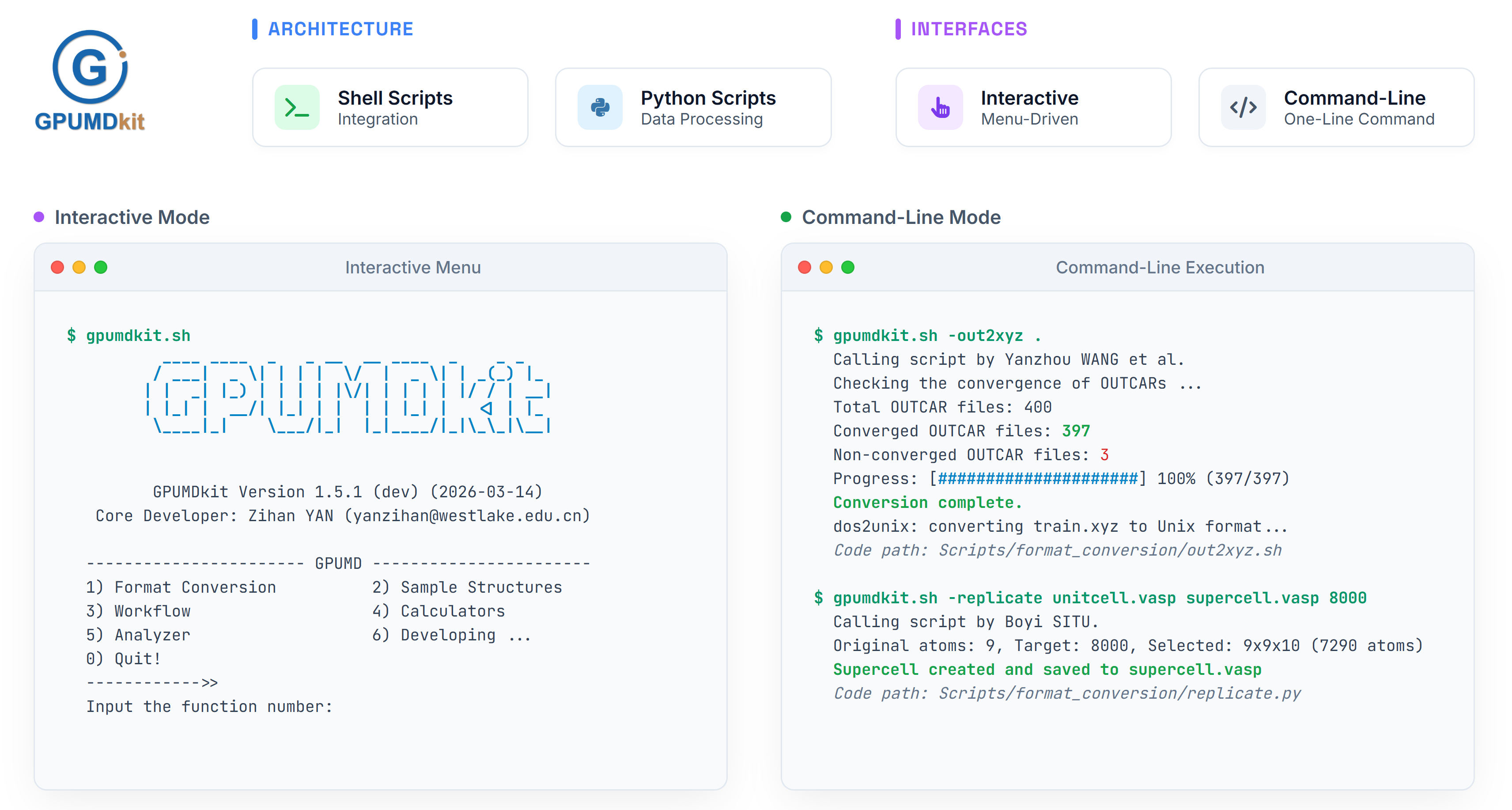}
  \caption{Schematic diagram of GPUMDkit.}
  \label{fig:schematic}
\end{figure*}

GPUMDkit is primarily written in Shell (for script orchestration and command-line integration) and Python (for data processing and analysis). It leverages established Python libraries, including ASE~\cite{larsen2017atomic}, dpdata~\cite{zeng2025dpdata}, pymatgen~\cite{ong2013python}, calorine~\cite{lindgren2024calorine}, and NepTrain~\cite{chen2025neptrain}, ensuring robustness and compatibility with existing frameworks. This modular design allows for easy maintenance and straightforward addition of new features without altering the core structure. The key functionalities of GPUMDkit are illustrated in Fig.~\ref{fig:schematic}.

Inspired by popular VASP post-processing toolkits such as VASPKIT~\cite{wang2021vaspkit,geng2025empowering} and qvasp~\cite{yi2020qvasp}, GPUMDkit provides two interface options to meet diverse user needs. The interactive mode delivers intuitive, step-by-step prompts, making it ideal for newcomers or those exploring complex workflows. The command-line interface, by contrast, caters to experienced users who require efficient execution and seamless integration into automated pipelines.

\subsection{Interactive Mode}
GPUMDkit provides a user-friendly interactive interface launched via the \verb'gpumdkit.sh' command. Users are presented with a menu-driven interface (Fig.~\ref{fig:schematic}) that organizes functionalities into clearly numbered categories, including format conversion, structure sampling, workflow management, calculators, and analysis tools. This hierarchical design allows users to navigate and execute a wide range of tasks through simple numeric selections.

\vspace{0.1cm}
\textbf{Format Conversion:} GPUMDkit supports conversion between a variety of common structure and trajectory formats, including input and output files from VASP~\cite{Kresse1994,Kresse1996}, CP2K~\cite{kuhne2020cp2k}, ABACUS~\cite{zhou2025abacus}, LAMMPS~\cite{thompson2022lammps}, and others. Both single structure files and multi-step trajectory files are supported, making this functionality particularly useful for post-processing MD simulations and preparing datasets for NEP training.
\vspace{0.1cm}

\textbf{Sample Structures:} The structure sampling module supports three main strategies: random sampling for rapid dataset construction, equally spaced sampling for uniform coverage of time series, and descriptor-based farthest point sampling (FPS) for optimal coverage of configuration space. FPS is particularly valuable for identifying representative atomic configurations while minimizing structural redundancy, thereby improving the training efficiency and generalization of the NEP model. Additionally, the module provides a random perturbation function that applies random atomic displacements to an initial structure, generating diverse training samples that are especially useful for highly symmetric systems or those with limited configuration space.
\vspace{0.1cm}

\textbf{Workflow Automation:} Although the NepTrain package~\cite{chen2025neptrain} already provides automated tools for NEP development, GPUMDkit offers a complementary and flexible workflow option. The module supports both fully automated and semi-automated (step-by-step) active learning cycles, encompassing iterative structure selection, DFT job preparation, model training, and validation. A key advantage is the fine-grained control it affords at each stage: users can pause the process to inspect sampled structures or adjust training settings before proceeding to the next iteration. This balance between automation and manual oversight helps users build high-quality NEP models while maintaining full transparency over the development process.
\vspace{0.1cm}

\textbf{Property Calculations:} The calculator module provides essential post-processing capabilities for GPUMD output analysis. Currently supported properties include radial distribution functions, self-diffusion coefficients, ionic conductivity calculations, density of atomistic states, and nudged elastic band pathway analysis. Additional functionality is under active development.
\vspace{0.1cm}

\textbf{Visualization and Analysis:} The analyzer module provides comprehensive data quality control capabilities, including training set composition analysis, charge balance verification, and outlier detection. Beyond these analysis functions, GPUMDkit also integrates a rich set of plotting tools for a wide range of visualization tasks, such as monitoring training progress, thermodynamic analysis, error assessment, and structural property visualization including radial distribution function, mean square displacement (MSD), and descriptor distributions.

\subsection{Command-Line Interface}
For users familiar with GPUMDkit, the command-line mode enables faster execution by directly passing arguments, supporting batch processing and seamless integration with automated computation pipelines. Here we briefly introduce the \verb'-plt' (visualization) and \verb'-time' (real-time monitoring) commands.

\vspace{0.1cm}
\textbf{Visualization Tools:} The \verb'-plt' command exemplifies the efficiency of the command-line interface. For instance, executing \verb'gpumdkit.sh -plt thermo' automatically invokes the corresponding visualization script to parse the \verb'thermo.out' output file from GPUMD, generating a comprehensive overview of thermodynamic quantities including temperature, pressure, potential and kinetic energy, lattice parameters, volume, and lattice angles (Fig.~\ref{fig:thermo}). This is particularly useful for assessing thermodynamic equilibrium, detecting phase transitions, or identifying simulation instabilities. A rich set of additional plotting scripts is accessible via the \verb'Scripts/plt_scripts' directory or the \verb'gpumdkit.sh -plt [options]' command, as summarized in Fig.~S1 of the Supporting Information (SI). These scripts automatically detect GUI availability and either display an interactive plot or save the output in PNG format. 

\vspace{0.1cm}
\textbf{Real-time Monitoring Features:} The real-time monitoring feature (\verb'gpumdkit.sh -time nep/gpumd') intelligently parses \verb'loss.out' or \verb'neighbor.out' files to dynamically assess computational progress and provide accurate completion time predictions. This function continuously tracks the execution efficiency of the NEP and GPUMD programs, allowing users to plan their time accordingly.

\vspace{0.1cm}
\textbf{Other Features:} Beyond the functions described above, GPUMDkit provides a range of additional command-line utilities covering common tasks and specialized computing requirements (see Fig.~S2 in SI).

\section{Case Studies: Phase Transition and Ionic Diffusion of \NoCaseChange{Li$_7$La$_3$Zr$_2$O$_{12}$}}

\begin{figure*}[htb]
  \centering
  \includegraphics[width=0.7\linewidth]{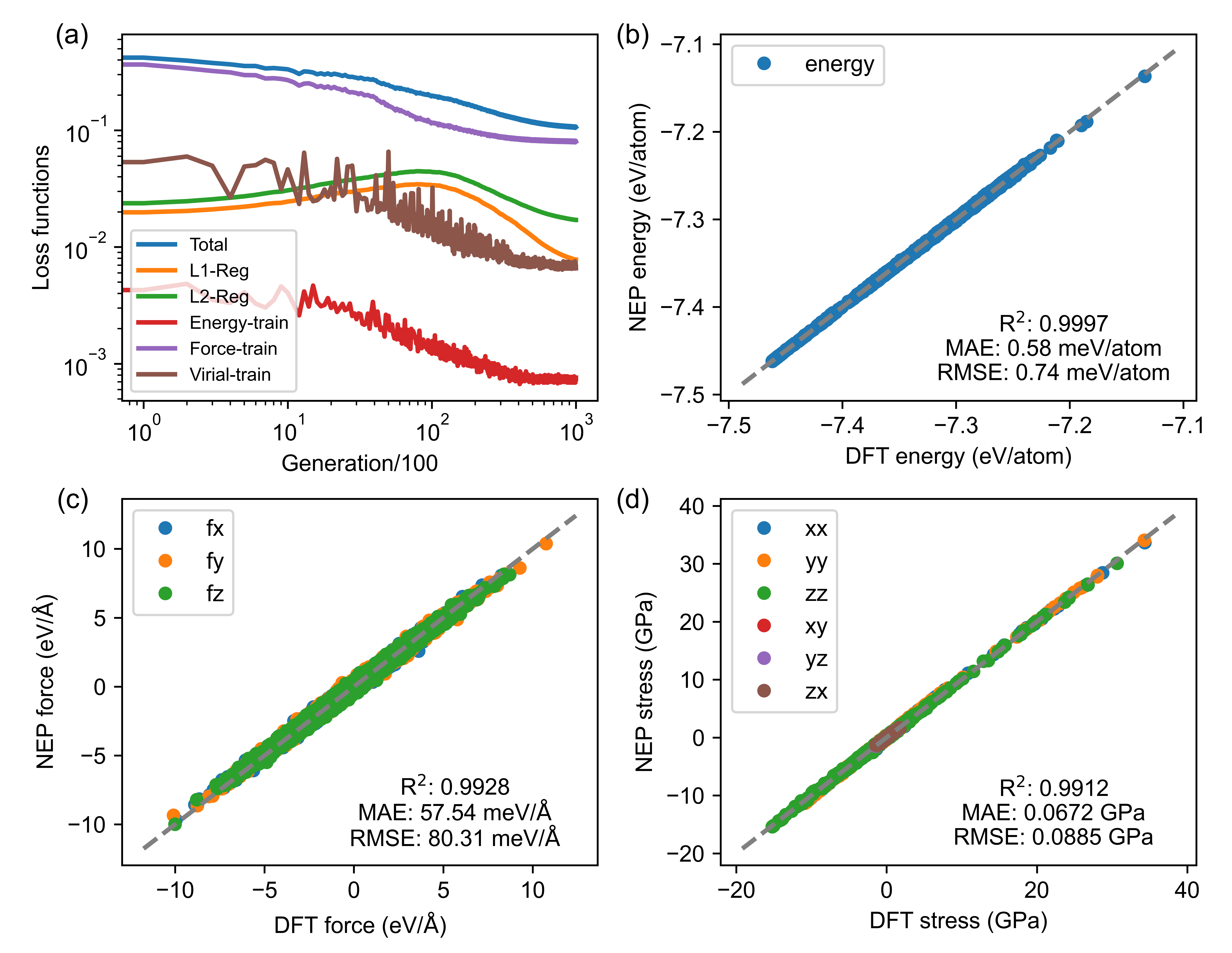}
  \caption{(a) Evolution of various terms in the loss function during the NEP training process. (b) Energy, (c) force, and (d) stress values from the NEP model, in comparison to the DFT reference data.}
  \label{fig:train}
\end{figure*}

Garnet-type \ce{Li7La3Zr2O_{12}} (LLZO) is one of the most promising solid electrolyte materials for next-generation all-solid-state lithium batteries, combining high ionic conductivity with excellent chemical and electrochemical stability against lithium metal~\cite{han2016electrochemical}. LLZO undergoes a phase transition at approximately \qty{900}{\kelvin}, from a low-temperature tetragonal phase (\textit{t}-LLZO) to a high-temperature cubic phase (\textit{c}-LLZO), accompanied by a 2--3 order-of-magnitude increase in ionic conductivity. This transition involves a rearrangement of the Li-ion sublattice, from fully occupied, ordered sites in \textit{t}-LLZO to partially occupied, disordered arrangements in \textit{c}-LLZO, fundamentally altering the Li-ion diffusion energy landscape and lowering the activation energy from \qty{1.227}{\eV} to \qty{0.309}{\eV}~\cite{yan2024impact}. The LLZO system therefore serves as an ideal testbed for demonstrating the capabilities of GPUMDkit in analyzing structural transitions and ion transport properties.

A NEP model was developed based on the training dataset from Yan and Zhu~\cite{yan2024impact}, which covers structural configurations representative of both tetragonal and cubic phases, including thermal snapshots at various temperatures, mechanically strained structures, and non-stoichiometric configurations with lithium-oxygen Schottky defect pairs. This diversity ensures robust model performance across a wide range of thermodynamic conditions. The training process and model accuracy were evaluated using the \verb'gpumdkit.sh -plt train' command, which visualizes the evolution of loss terms and generates parity plots of energies, forces, and stresses against DFT reference values (Fig.~\ref{fig:train}). The energy predictions exhibit excellent agreement ($R^2$ \textgreater\ 0.999) with an RMSE of \qty{0.74}{meV/atom}, force predictions achieve an root mean square error (RMSE) of \qty{80.31}{\meV/\angstrom}, and stress tensor predictions yield an RMSE of \qty{0.0885}{\GPa}, confirming the accuracy and reliability of the model.

\begin{figure*}[htb]
  \centering
  \includegraphics[width=0.9\linewidth]{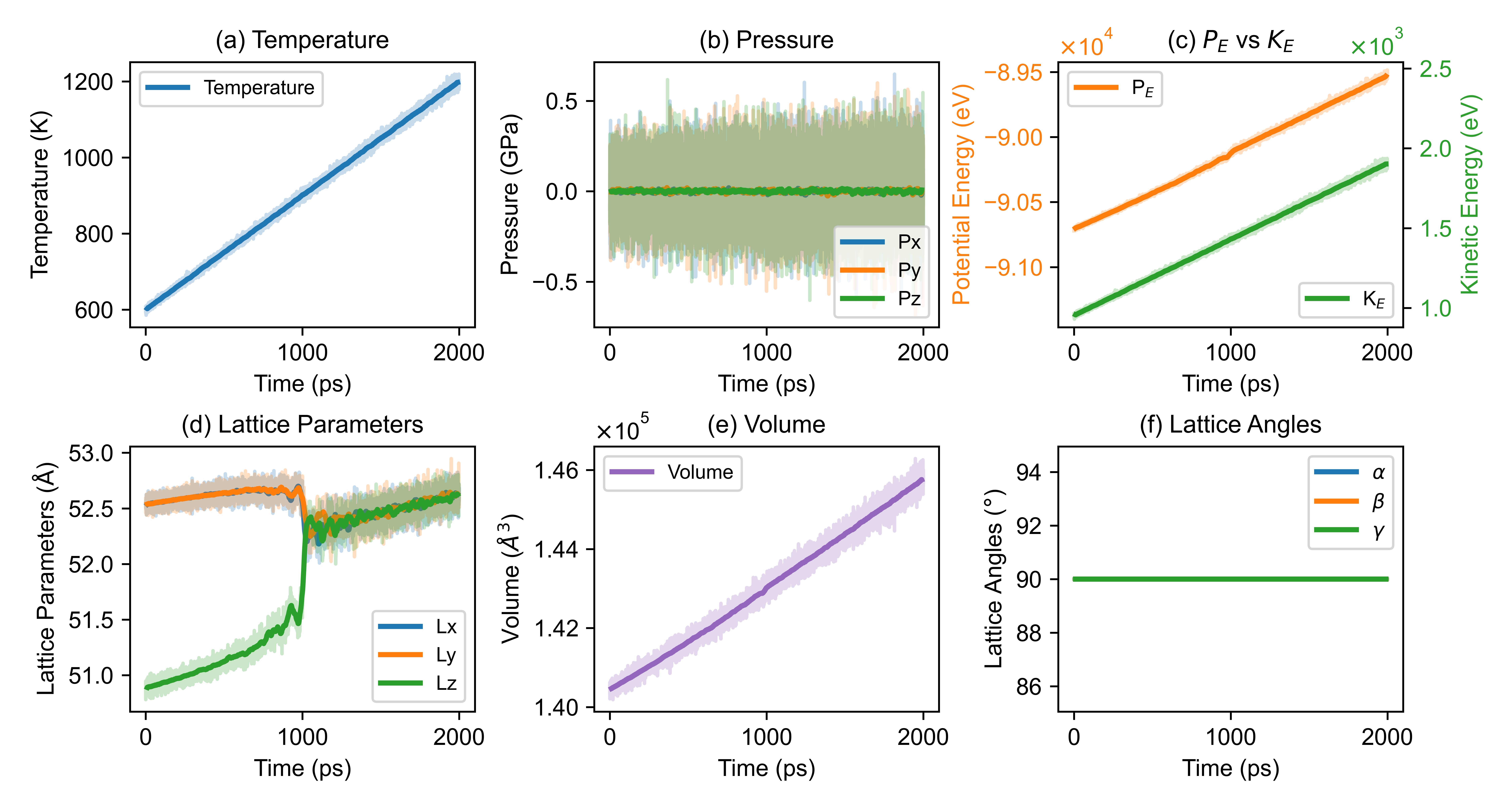}
  \caption{Thermodynamic properties of LLZO during heating from \qty{600}{\kelvin} to \qty{1200}{\kelvin}: (a) temperature, (b) pressure, (c) potential and kinetic energy, (d) lattice parameters, (e) volume, and (f) lattice angles.}
  \label{fig:thermo}
\end{figure*}

To investigate the tetragonal-to-cubic phase transition in LLZO, we performed MLMD simulations using GPUMD with the NEP model. The simulations employed a $4\times4\times4$ supercell containing 12,288 atoms under the isothermal-isobaric (NPT) ensemble with Martyna-Tuckerman-Tobias-Klein integrators and a timestep of \qty{1}{{\femto\second}}. The system was heated from \qty{600}{\kelvin} to \qty{1200}{\kelvin} at a rate of \qty{300}{\kelvin\per\nano\second}, sufficient to capture the phase transition, as verified in the previous work~\cite{yan2025improving}. Thermodynamic properties throughout the heating process were monitored using the \verb'gpumdkit.sh -plt thermo' command, which tracks temperature, pressure, potential and kinetic energy, lattice parameters, volume, and lattice angles, as shown in Fig.~\ref{fig:thermo}.

The tetragonal-to-cubic phase transition is reflected in the evolution of the lattice parameters (Fig.~\ref{fig:thermo}(d)). Below \qty{900}{\kelvin}, all three parameters exhibit normal thermal expansion while maintaining the tetragonal relationship (\textit{a} = \textit{b} \textgreater\ \textit{c}). At approximately \qty{900}{\kelvin}, a dramatic structural change occurs: \textit{a} and \textit{b} contract while \textit{c} expands until all three converge to equivalent values (\textit{a} = \textit{b} = \textit{c}), characteristic of the cubic phase. Above this temperature, the three parameters resume thermal expansion as equivalent cubic lattice constants. This transition temperature agrees well with previous experimental and computational reports~\cite{yan2024impact,chen2015study}, validating the ability of our NEP model to capture the phase transition thermodynamics.

Notably, the volume and potential energy also exhibit subtle but discernible anomalies near the transition point, consistent with the first-order nature of the tetragonal-to-cubic phase transition in LLZO. These changes are less pronounced than those in the lattice parameters, as the transition is primarily driven by an order-disorder transition of the Li-ion sublattice rather than by significant structural changes in the rigid La-Zr-O framework. This order-disorder transition reshapes the Li-ion diffusion energy landscape, significantly enhancing the ionic transport of LLZO and giving rise to the dramatic conductivity increase observed in the cubic phase~\cite{yan2024impact}.

\begin{figure*}[htb]
  \centering
  \includegraphics[width=0.99\linewidth]{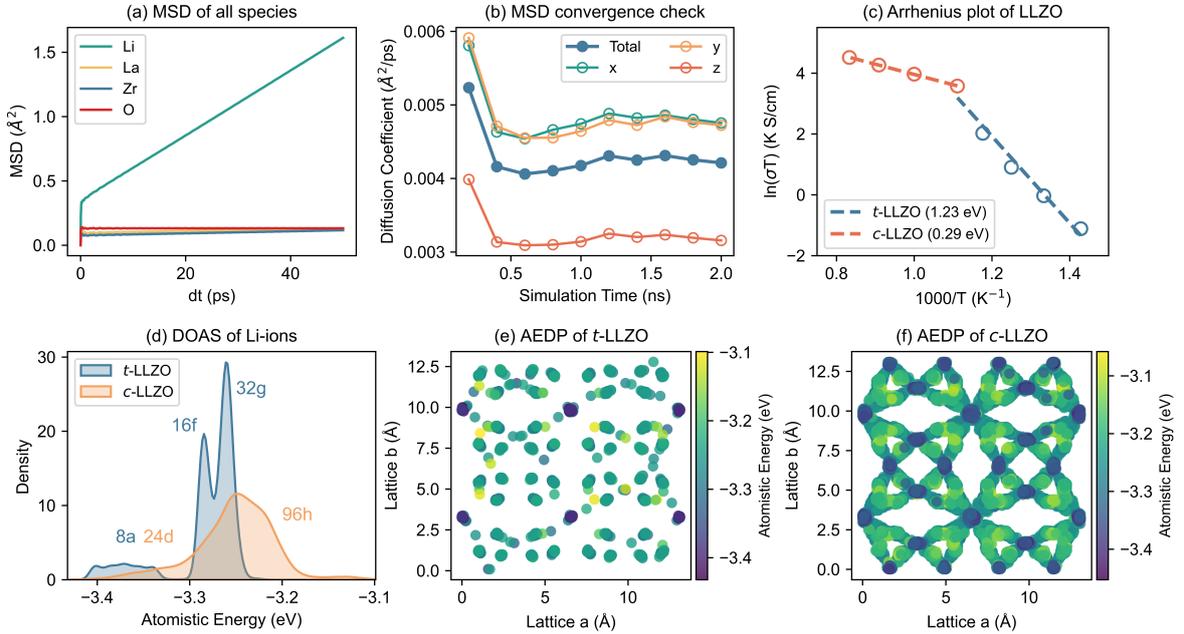}
  \caption{(a) The mean square displacement (MSD) of each atomic species at \qty{800}{\kelvin}. (b) The evolution of Li-ion diffusivities during the simulation. (c) Arrhenius plot of Li-ion diffusion in \textit{t}-LLZO and \textit{c}-LLZO. (d) Density of atomistic states of Li-ions in LLZO. Atomistic energy distribution plot (AEDP) of Li-ions in (e) \textit{t}-LLZO and (f) \textit{c}-LLZO.}
  \label{fig:garnet-analysis}
\end{figure*}

To quantify this effect, we performed \qty{2}{\nano\second} MLMD simulations in the temperature range of \qtyrange{700}{1200}{\kelvin}, using the same $4\times4\times4$ supercell. Fig.~\ref{fig:garnet-analysis}(a), generated using the \verb'gpumdkit.sh -plt msd_all' command, shows the MSD of each atomic species at \qty{800}{\kelvin}. Li-ions exhibit high mobility with a linear MSD, confirming long-range diffusion rather than local vibrations, while the framework atoms (La, Zr, O) remain essentially immobile, preserving the structural integrity of the La-Zr-O framework. Fig.~\ref{fig:garnet-analysis}(b), produced via \verb'gpumdkit.sh -plt msd_conv', tracks the evolution of Li-ion diffusion coefficients throughout the \qty{2}{\nano\second} simulation, which converge to stable values within the first \qty{1}{\nano\second}, indicating sufficient sampling for reliable statistics. Notably, the diffusion exhibits significant anisotropy at \qty{800}{\kelvin}: Li-ion mobility in the \textit{x} and \textit{y} directions is  higher than in the \textit{z} direction, reflecting the crystallographic constraints of the tetragonal structure that preferentially support ion transport within specific crystal planes.

The dramatic effect of the phase transition is evident in the Arrhenius plot (Fig.~\ref{fig:garnet-analysis}(c)), which reveals a large change in activation energies across the critical temperature. Li-ions diffusion in the \textit{c}-LLZO has a low activation energy (\qty{0.29}{\eV}), whereas in the \textit{t}-LLZO, the activation energy is significantly higher (\qty{1.23}{\eV}). These values align well with the previous computational results~\cite{yan2024impact}, validating the accuracy of the retrained NEP model in reproducing transport properties. The contrast in transport behavior suggests that while macroscopic thermodynamic properties exhibit only subtle changes during phase transitions (Fig.~\ref{fig:thermo}), the underlying structural change has a significant impact on Li-ion dynamics. The order-disorder transition alters the energetic landscape for Li-ions diffusion, transforming LLZO into the cubic phase as an excellent superionic material.

To understand the microscopic mechanisms underlying the order-disorder transition and its impact on Li-ions diffusion, we employ the density of atomistic states (DOAS) analysis implemented in GPUMDkit. This method was first proposed by Wang \textit{et al.} and successfully revealed frustration in superionic conductors~\cite{wang2023frustration}. The DOAS approach is based on the fundamental principle that in MLIPs, the total system energy comes from summing individual atomistic energies, where the atomistic energy of each atom is determined by its local atomic environment. By constructing the atomistic energy distribution of different atoms, DOAS provides direct insight into the distribution of local environments and their energy landscape.

Fig.~\ref{fig:garnet-analysis}(d) shows the Li-ion DOAS distributions for both \textit{t}-LLZO and \textit{c}-LLZO, revealing significant difference in atomistic energies among different Li-ion sites. In \textit{t}-LLZO, the Li-ion energy distribution displays three distinct, sharp peaks corresponding to three kinds of  fully occupied crystallographic sites: the tetrahedral 8\textit{a} position and two octahedral positions 16\textit{f} and 32\textit{g}. These peaks reflect the ordered Li-ion sublattice in the \textit{t}-LLZO. In contrast, the \textit{c}-LLZO displays a markedly different energy landscape. The Li-ion DOAS becomes much broader and more diffuse, with a wider energy distribution range and less distinct peak separation. The two main features correspond to partially occupied tetrahedral (24\textit{d}) and octahedral (96\textit{h}) sites, but their energy levels are much closer compared to the tetragonal phase. This energy flattening in the cubic phase directly explains the dramatic enhancement in ionic conductivity after the phase transition. The reduced energy barriers between different Li-ion sites create a more favorable landscape for Li-ions diffusion~\cite{wang2023frustration}. Therefore, DOAS analysis provides quantitative evidence for an order-disorder transition mechanism, where the discrete energy landscape of Li-ions in the \textit{t}-LLZO is transformed into a more continuous energy landscape in the \textit{c}-LLZO.

To further demonstrate the Li-ion sublattice rearrangement, Figs.~\ref{fig:garnet-analysis}(e,f) show the atomistic energy distribution plots (AEDP), depicting the spatial distribution of Li-ion energy states in both phases. In \textit{t}-LLZO, Li-ions exhibit a highly ordered spatial distribution with clear energy segregation. Low-energy Li-ions (deep purple, \qty{-3.45}{\eV}) corresponding to the 8\textit{a} tetrahedral sites form a regular and well-defined pattern, while mid-energy Li-ions (cyan, \qty{-3.3}{} to \qty{-3.15}{\eV}) are systematically arranged near well-defined crystallographic sites (16\textit{f} and 32\textit{g}). Furthermore, AEDP reveals the presence of high-energy Li-ions (bright yellow, \textgreater\ \qty{-3.15}{\eV}), although these states are minimally represented as determined by DOAS. In contrast, the cubic phase displays a markedly different energy distribution. The Li-ions are distributed more connected along diffusion channels, and the discrete energy landscape observed in the \textit{t}-LLZO is no longer present, replaced by a more continuous energy spread. Notably, the number of mid- and high-energy Li-ions increases significantly, indicating a broader occupation across energy states that facilitates ion transport. This spatial analysis clearly illustrates the order-disorder transition mechanism in LLZO: the changed energy landscape enables Li-ions to access a wider range of positions that were previously prohibited, creating a network of available sites and pathways for ion transport. This smoothed energy distribution replaces the distinct energy barriers that restrict Li-ions diffusion in the \textit{t}-LLZO, leading to the excellent ionic conductivity of cubic LLZO.

\section{Case Studies: Phase Transition and Topological Structure in \NoCaseChange{(Pb,Sr)TiO$_3$}}

Ferroelectric materials, characterized by a spontaneous polarization that can be reversed by an external electric field, are fundamental to a broad spectrum of functional devices, including nonvolatile memory, sensors, and piezoelectric actuators~\cite{Lines77,Damjanovic01p191,Garcia14p4289,Li25peadn4926}. Among these, \ce{ABO3}-type perovskite oxides serve as a technologically vital class, where the structural flexibility provided by the distinct ionic radii of the A-site and B-site cations allows for the precise tuning of functional properties. An important example is \ce{PbTiO3} (PTO), a prototypical ferroelectric that exhibits a robust tetragonal ground state with a large spontaneous polarization and a high Curie temperature ($T_\text{C}$) of \qty{765}{\kelvin}~\cite{Shirane51p265}. In contrast, \ce{SrTiO3} (STO) behaves as a quantum paraelectric; while it typically remains nonpolar, ferroelectricity can be induced through external stimuli such as epitaxial strain or terahertz pulses~\cite{Xu20p3141,Li19p1079}. Furthermore, \ce{PbTiO3}/\ce{SrTiO3} superlattices have recently emerged as a versatile platform for exploring emergent topological dipolar structures—including flux closures, vortices, and skyrmions—which arise from the complex interplay of electric, elastic, and gradient energies at the nanoscale~\cite{Tang15p547,Yadav16p198,Das19p368,jiyuan24p13}. 
In this section, we employ GPUMDkit to perform the analysis and visualization of these important materials.

To accurately model the atomistic dynamics of this system, we developed a NEP model covering \ce{PbTiO3}, \ce{SrTiO3}, and their solid solutions. The training dataset was derived from our previous study, which utilized a modular development strategy to efficiently sample the configuration space~\cite{Wu23p144102}. The dataset comprehensively covers the end-members, including the tetragonal $P4mm$ and cubic $Pm\bar{3}m$ phases of \ce{PbTiO3}, as well as the cubic and antiferrodistortive $I4/mcm$ phases of \ce{SrTiO3}. It also incorporates random solid solution configurations (\ce{Pb_xSr_{1-x}TiO3}) generated through extensive sampling with molecular dynamics and Monte Carlo swapping. The robust fitting performance of the developed NEP model was confirmed by the parity plots of energy, atomic forces, and stress tensors, which were automatically generated using the \verb'gpumdkit.sh -plt train_test' command (Fig. S3). The model achieves RMSE of \qty{1.99}{meV/atom} for energy and \qty{89.11}{\meV\per\angstrom} for atomic forces, and \qty{0.217}{\GPa} for stress tensors on the training dataset, demonstrating high fidelity to the reference DFT data.

\begin{figure*}[]
  \centering
  \includegraphics[width=0.95\linewidth]{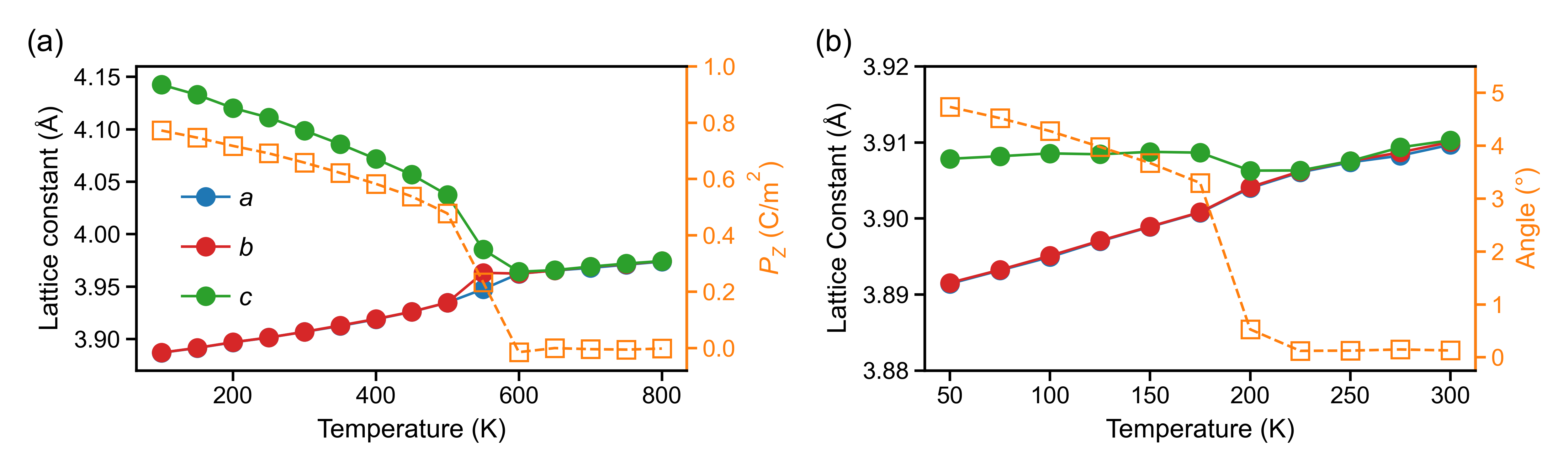}
  \caption{
    Temperature dependence of structural properties in bulk \ce{PbTiO3} and \ce{SrTiO3}. 
    (a) Lattice constants (left axis) and spontaneous polarization along the polar axis $P_z$ (right axis) of \ce{PbTiO3} as a function of temperature, showing the ferroelectric tetragonal ($P4mm$) to paraelectric cubic ($Pm\bar{3}m$) transition at $T_\text{C} \approx \qty{600}{\kelvin}$.
    (b) Lattice constants (left axis) and the \ce{TiO6} octahedral tilt angle (right axis) of \ce{SrTiO3}, illustrating the antiferrodistortive phase transition at $T_\text{C} \approx \qty{225}{\kelvin}$.
    }
  \label{fig:fe_bulk}
\end{figure*}

We initiate our analysis by investigating the temperature-driven phase transitions in bulk \ce{PbTiO3} and \ce{SrTiO3}. By utilizing the \verb'gpumdkit.sh -calc avg-stru' command, we efficiently extracted the ensemble-averaged structures from the MD trajectories for detailed analysis. 
The results are presented in Fig.~\ref{fig:fe_bulk}. For \ce{PbTiO3} (Fig.~\ref{fig:fe_bulk}(a)), we observe a distinct phase transition from the ferroelectric tetragonal phase to paraelectric cubic phase at a $T_\text{C}$ of approximately \qty{600}{\kelvin}. This result is in good agreement with previous theoretical studies using similar potentials, though slightly lower than the experimental value of \qty{765}{\kelvin}~\cite{Shirane51p265}. To quantify the strength of polarization, we define the local polarization $\mathbf{P}^m (t)$ for a unit cell $m$ using the formalism~\cite{Qi16p134308}:

\begin{equation}
    \mathbf{P}^m (t) = \frac{1}{V_{\mathrm{u.c.}}} \left[\frac{1}{8} Z_\mathrm{A}^* \sum_{k=1}^8 \mathbf{r}_{\mathrm{A},k}^m (t) + Z_{\mathrm{B}}^* \mathbf{r}_\mathrm{B}^m (t) + \frac{1}{2} Z_{\mathrm{O}}^* \sum_{k=1}^6 \mathbf{r}_{\mathrm{O},k}^m (t)\right]
\end{equation}
where $V_{\mathrm{u.c.}}$ is the unit cell volume, $\mathbf{r}$ represents the atomic positions and $Z$ represents the Born effective charges. We computed this quantity using the \verb'gpumdkit.sh -calc pol-abo3' command. The resulting spontaneous polarization, depicted as the orange line in Fig.~\ref{fig:fe_bulk}(a), remains robust at low temperatures and vanishes at $T_\text{C}$ as the lattice constants $a$ and $c$ converge to the cubic phase.
Similarly, for \ce{SrTiO3}, a structural phase transition is observed at approximately \qty{225}{\kelvin} (Fig.~\ref{fig:fe_bulk}(b)), consistent with previous MD simulations~\cite{Wu23p144102}. While this theoretical $T_\text{C}$ is higher than the experimental value of \qty{105}{\kelvin}, it is well-understood that \ce{SrTiO3} is a quantum paraelectric; incorporating nuclear quantum effects into the simulation (e.g., via path-integral MD) has been shown to suppress the transition temperature, bringing it closer to the experimental result~\cite{Wu22p224102}. The transition is characterized by the antiferrodistortive rotation of the oxygen octahedra. We extracted the associated order parameter using the \verb'gpumdkit.sh -calc oct-tilt' command; the calculated \ce{TiO6} tilt angle (orange line, Fig.~\ref{fig:fe_bulk}(b)) diminishes with heating and disappears at the transition temperature. These results confirm that the trained NEP model faithfully reproduces the thermodynamic behavior of the pure bulk phases.


\begin{figure*}[]
  \centering
  \includegraphics[width=0.95\linewidth]{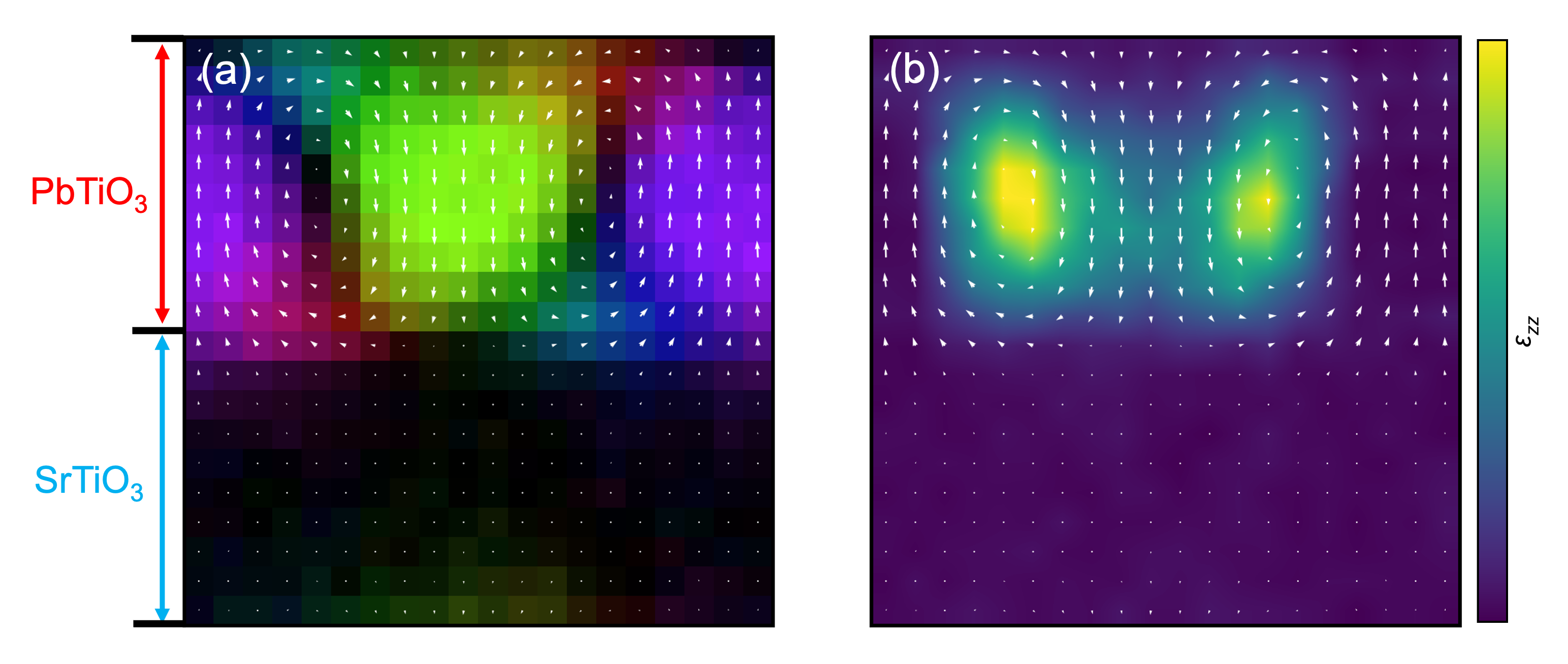}
  \caption{Atomistic modeling of topological polar structures in the (PbTiO$_3$)$_{10}$/(SrTiO$_3$)$_{10}$ superlattice at \qty{300}{\kelvin}. (a) Side view of the local polarization distribution in the $x$-$z$ plane, revealing the formation of polar vortex arrays within the ferroelectric \ce{PbTiO3} layers. The arrows indicate the direction and magnitude of the local dipoles. (b) Real-space map of the local out-of-plane dielectric permittivity $\varepsilon_{zz}$ calculated using the fluctuation-dissipation theorem. The map reveals enhanced dielectric susceptibility at the vortex cores and domain walls compared to the domain regions.}
  \label{fig:fe_sl}
\end{figure*}

Subsequently, we modeled the (\ce{PbTiO3})$_{10}$/(\ce{SrTiO3})$_{10}$ superlattice to investigate its topological polar structures. A supercell comprising alternating 10-unit-cell layers of PTO and STO was constructed, and MD simulations were performed in the NPT ensemble to equilibrate the system. The resulting local polarization distribution, visualized in Fig.~\ref{fig:fe_sl}(a), exhibits a distinct polar vortex array character, reflecting the delicate balance between depolarization fields and domain wall energies in the confined ferroelectric layers. These topological textures are in good agreement with previous theoretical results~\cite{jiyuan24p13}. To probe the local dielectric properties, we calculated the local out-of-plane dielectric permittivity $\varepsilon_{zz}^m$ for the unit cell $m$, defined following the fluctuation-based approach~\cite{Liu16p094102}:
\begin{equation}
\varepsilon_{ij}^{m} = \frac{V_{\mathrm{u.c.}}}{\varepsilon_0 k_\text{B} T} \left(\langle P_i^m  P_j^m\rangle - \langle P_i^m \rangle \langle P_j^m \rangle\right)   
\end{equation}
where $V_{\mathrm{u.c.}}$ is the volume of the unit cell, $k_\text{B}$ is the Boltzmann constant, $T$ is the temperature, $\varepsilon_{0}$ is the vacuum permittivity, and $P_{i}^{m}$ is the instantaneous local polarization of the unit cell $m$ along the $i$ direction. This value quantifies the local polarization fluctuations. As shown in Fig.~\ref{fig:fe_sl}(b), the permittivity map reveals a significant enhancement at the vortex cores and domain walls compared to the domain regions, indicating that these topologically frustrated regions serve as highly susceptible ``hotspots" for dielectric response.

\section{Case Study: Thermal transport of graphene}

GPUMD has become a widely used platform for phonon-mediated thermal-transport simulations, driven by two main advantages: (i) GPU acceleration combined with the high accuracy enabled by NEP framework, and (ii) comprehensive coverage of major thermal-transport formalisms within a unified workflow. Building on standard GPUMD outputs, GPUMDkit provides an end-to-end post-processing pipeline that converts raw trajectories and numerical files into publication-ready thermal-transport results through standardized analysis, consistent visualization, and automated uncertainty estimation. In particular, GPUMDkit simplifies common but error-prone tasks such as parsing outputs, aggregating independent runs, checking convergence, computing key correlation functions and transport properties, and producing standardized plots and tables.

As a representative high-thermal-conductivity two-dimensional material, monolayer graphene is used here to demonstrate GPUMDkit’s capability in processing and visualizing simulations performed using equilibrium molecular dynamics (EMD), non-equilibrium molecular dynamics (NEMD), and homogeneous non-equilibrium molecular dynamics (HNEMD). The main text focuses on HNEMD as the primary example, while the EMD and NEMD visualizations are provided in the SI (Figs.~S4--5).

Based on linear-response theory, the HNEMD method mimics the effect of a thermal gradient by applying a directional external driving force $\mathbf{F}_{i}^{\mathrm{ext}}$ to each atom $i$~\cite{evans1982homogeneous-c1b,fan2019homogeneous-6b0}, defined as:
\begin{equation}
\label{equation:fe}
\mathbf{F}_{i}^{\mathrm{ext}} = \mathbf{F}_{\mathrm{e}} \cdot \mathbf{W}_{i},
\end{equation}
where $\mathbf{F}_\mathrm{e}$ is a small vector field with dimension of inverse length, and $\mathbf{W}_{i}$ is the per-atom virial tensor. The resulting heat current $\mathbf{J}=\sum_i \mathbf{W}_i\cdot \mathbf{v}_i$ yields a non-equilibrium ensemble average $\langle \boldsymbol{J}(t)\rangle_{\mathrm{ne}}$, from which the running thermal conductivity is obtained as:
\begin{equation}
\label{equation:kappa}
\kappa(t)=\frac{\langle \mathbf{J}(t)\rangle_{\mathrm{ne}}}{T V \mathbf{F}_{\mathrm{e}}},
\end{equation}
where $T$ is the temperature, $V$ is the volume, and $\mathbf{v}_{i}$ is the velocity of atom $i$. To assess convergence, $\kappa(t)$ is further expressed as a cumulative average, $\kappa(t)=\frac{1}{t}\int_0^t \kappa(\tau)\mathrm{d}\tau$.

\begin{figure*}[b]
	\centering
	\includegraphics[width=0.7\linewidth]{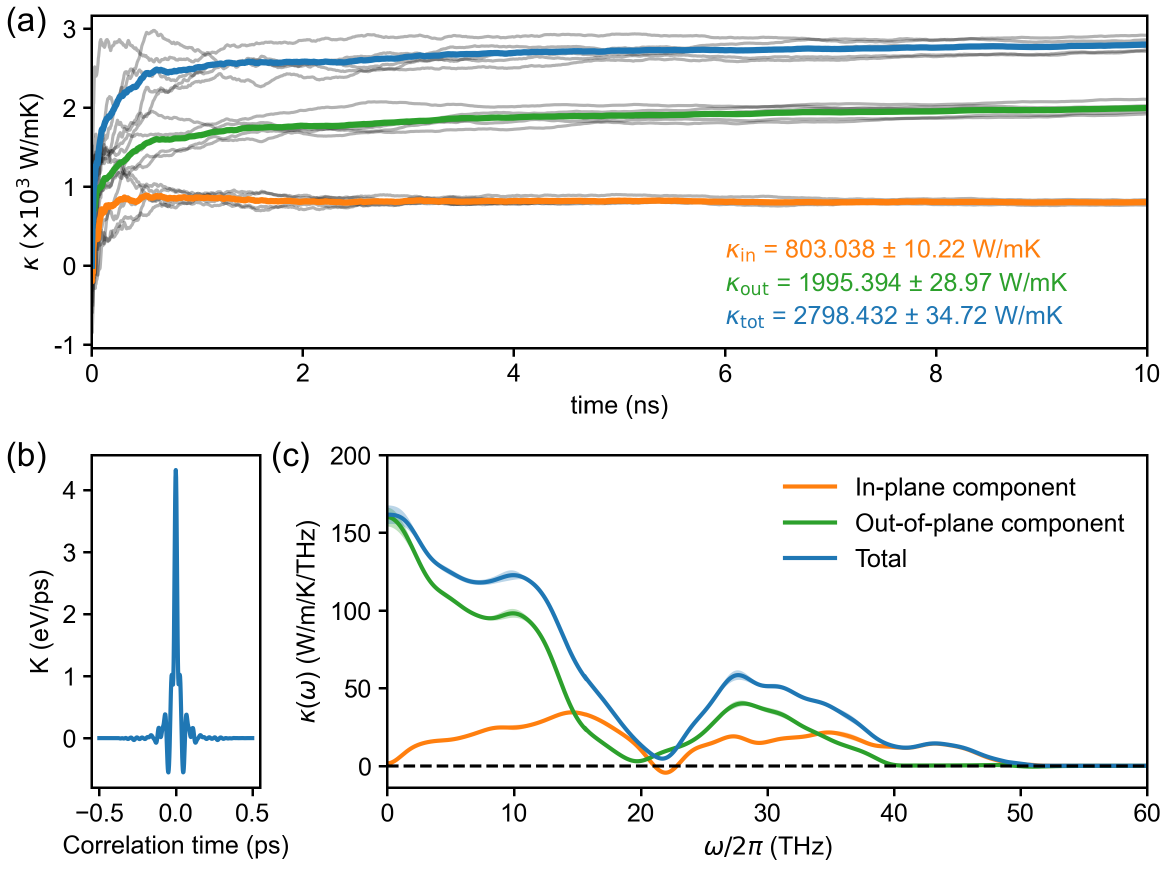}
	\caption{Thermal transport of monolayer graphene by HNEMD at \qty{300}{\kelvin}. (a) Running thermal conductivity as a function of simulation time, decomposed into in-plane and out-of-plane phonon contributions. (b) Virial-velocity correlation function versus time. (c) Spectral thermal conductivity with the same in-plane and out-of-plane decomposition. The results are from five independent simulations, with error bars shown by standard errors.}
	\label{fig:heat}
\end{figure*}

GPUMDkit enables rapid analysis and figure generation for HNEMD simulations via the command \verb"gpumdkit.sh -plt hnemd" (Fig.~\ref{fig:heat}). The script automatically detects whether spectral heat-current decomposition was performed. If spectral data are available, GPUMDkit outputs and visualizes the corresponding spectral thermal conductivity; otherwise, it reports only the standard HNEMD-based running thermal conductivity. For spectral post-processing, users may additionally specify a cut-off frequency to define the decomposition range. 

Fig.~\ref{fig:heat}(a) shows the time evolution of the thermal conductivity of monolayer graphene and the contribution from the in-plane and out-of-plane phonons at \qty{300}{\kelvin}. The system reaches a clear steady-state plateau within the first \qty{2}{\nano\second}, indicating rapid convergence of the heat-current response under the applied driving force. The converged thermal conductivity contributed from the in-plane phonons, $\kappa_{\mathrm{in}} \approx 803~\mathrm{W\,m^{-1}\,K^{-1}}$, is significantly smaller than the out-of-plane contribution, $\kappa_{\mathrm{out}} \approx 1995~\mathrm{W\,m^{-1}\,K^{-1}}$, resulting in a total thermal conductivity of $\kappa_{\mathrm{tot}} \approx 2798~\mathrm{W\,m^{-1}\,K^{-1}}$. It reflects the intrinsic phonon-transport characteristics of graphene, where heat conduction is dominated by out-of-plane phonon modes with long mean free paths, arising predominantly from flexural phonons that are efficiently excited under non-equilibrium driving.

The Virial-velocity autocorrelation function shown in Fig.~\ref{fig:heat}~(b) exhibits a sharp central peak with rapidly decaying oscillations, suggesting strong short-time correlations and weak long-time memory effects. This behaviour is characteristic of crystalline graphene, where phonon scattering is limited and momentum-conserving normal processes play a significant role at room temperature.

The spectral decomposition of thermal conductivity in Fig.~\ref{fig:heat}~(c) further reveals the frequency-resolved contributions of different phonon modes. Low-frequency phonons dominate the thermal conductivity, consistent with the long lifetimes and large group velocities of acoustic phonons. In particular, the out-of-plane contribution shows substantial weight in the low- and mid-frequency range, highlighting the important role of flexural phonons in thermal transport, despite their quadratic dispersion. At higher frequencies ($>40~$THz), the spectral thermal conductivity decays rapidly, indicating limited contributions from optical phonons due to their low group velocities and enhanced scattering.



For quick access to detailed parameter definitions and usage options, GPUMDkit offers a built-in help interface, e.g., \verb"gpumdkit.sh -plt hnemd -h". The same help mechanism applies to EMD and NEMD by replacing \verb"hnemd" with the corresponding method name, allowing users to efficiently identify available arguments and their functions.

\section{Conclusions}

In this work, we introduced GPUMDkit, a comprehensive toolkit for GPUMD and NEP programs. By integrating a wide range of functionalities into a unified interface, including format conversion, structure sampling, automated workflows, property calculators, and visualization tools, GPUMDkit enables users to move rapidly from data preparation to property analysis without extensive custom coding. Both interactive and command-line interfaces are provided to accommodate users with different levels of programming experience. The modular architecture of GPUMDkit ensures straightforward maintenance and extensibility, allowing new features to be incorporated without altering the core structure. Overall, GPUMDkit substantially lowers the barrier to entry for GPUMD and NEP simulations and enhances the productivity of researchers across diverse application domains.

The capabilities of GPUMDkit were illustrated through three representative case studies: the order-disorder phase transition and ionic transport in the LLZO solid electrolyte, the structural phase transitions and topological polar structures in ferroelectric \ce{(Pb,Sr)TiO3} systems, and the phonon-mediated thermal transport in monolayer graphene. These examples collectively demonstrate the ability of GPUMDkit to connect macroscopic observations with atomic-scale mechanisms, enabling deeper understanding of complex material behavior.

\section*{Acknowledgements}

Z. Yan and Y. Zhu acknowledge the support from the National Natural Science Foundation of China (No. 22509162 and 225B2917). Z. Fan acknowledges the support from the Science Foundation from Education Department of Liaoning Province (No. LJ232510167001). X. Wu acknowledges the support from JSPS through a Postdoctoral Fellowship for Research in Japan (P24058) and Grants-in-Aid for Scientific Research (No. 24KF0027). S. Liu acknowledges the support from Zhejiang Provincial Natural Science Foundation of China (LR25A040004).

\section*{Data Availability}
Source files of case studies will be available on the GitHub repository at \url{https://github.com/zhyan0603/SourceFiles}.

\section*{Code Availability}
The source code of GPUMDkit is freely available at \url{https://github.com/zhyan0603/GPUMDkit}. Documentations can be found at \url{https://zhyan0603.github.io/GPUMDkit}.

\section*{Competing Interests}
The authors declare no competing interests.

\bibliography{references}

\end{document}